\pdfoutput=1

\documentclass[11pt]{article}

\usepackage[]{acl}

\usepackage{times}
\usepackage{latexsym}

\usepackage[T1]{fontenc}

\usepackage[utf8]{inputenc}

\usepackage{microtype}

\usepackage{inconsolata}

\usepackage{graphicx} 
\usepackage{subcaption}
\usepackage{multirow}
\usepackage{booktabs}
\usepackage{placeins}

\usepackage{tikz}
\usetikzlibrary{calc}
\usepackage{hyperref}

\title{On the Semantic Latent Space of Diffusion-Based Text-to-Speech Models}

\author{Miri Varshavsky-Hassid$^*$ \quad Roy Hirsch$^*$ \quad Regev Cohen \quad Tomer Golany \\ \textbf{Daniel Freedman} \quad \textbf{Ehud Rivlin} \\ \\ \normalfont{Verily AI} \\ \normalfont{\texttt{\{mirivar, royhirsch, regevcohen\}@google.com}}}

\usepackage{amsmath}
\usepackage{amssymb}
\usepackage[inline]{enumitem}

\tikzset{AST style/.style={fill=none}}
\newcommand{\tikzmark}[2]{\tikz[overlay,remember picture,anchor=base] \node (#1) {#2};}
\newcommand{\ThreeAst}{\ensuremath{\boldsymbol{{\ast}{\ast}{\ast}}}}
\newcommand{\OneAst}{\ensuremath{\boldsymbol{\ast}}}

\def\rvh{{\mathbf{h}}}

\def\rvv{{\mathbf{v}}}
\def\rvw{{\mathbf{w}}}
\def\rvx{{\mathbf{x}}}

\begin{document}
\maketitle
\def\thefootnote{*}\footnotetext{Equal contribution}
\let\thefootnote\relax\footnote{Paper accepted to ACL 2024.}
\begin{abstract}
The incorporation of Denoising Diffusion Models (DDMs) in the Text-to-Speech (TTS) domain is rising, providing great value in synthesizing high quality speech. Although they exhibit impressive audio quality, the extent of their semantic capabilities is unknown, and controlling their synthesized speech's vocal properties remains a challenge. Inspired by recent advances in image synthesis, we explore the latent space of frozen TTS models, which is composed of the latent bottleneck activations of the DDM's denoiser. We identify that this space contains rich semantic information, and outline several novel methods for finding semantic directions within it, both supervised and unsupervised. We then demonstrate how these enable off-the-shelf audio editing, without any further training, architectural changes or data requirements. We present evidence of the semantic and acoustic qualities of the edited audio, and provide supplemental samples: \href{https://latent-analysis-grad-tts.github.io/speech-samples/}{https://latent-analysis-grad-tts.github.io/speech-samples/}.
\end{abstract}

\section{Introduction}
Denoising Diffusion Models (DDMs)~\cite{sohl2015deep} have emerged as a powerful generative tool across a broad variety of tasks and domains. In particular, Text-to-Speech (TTS) systems based on diffusion have shown high-quality speech generation capabilities~\cite{huang2022prodiff, shen2023naturalspeech}. Although these exhibit improved quality, the extent to which they capture semantic information is yet to be uncovered, and the ability to \textit{control} the vocal properties (e.g. volume, pitch, gender) of their generated speech is limited. Uncovering the semantic capabilities of TTS diffusion models will allow editing the properties of synthesized speech, which is essential in real-world applications, such as human-machine interaction.

Diffusion-based TTS methods, such as WaveGrad and Diff-Wave, condition the generation process on mel-spectogram input \citep{chen2020wavegrad, kong2020diffwave}. More recent advances such as Diff-TTS, WaveGrad2, and Grad-TTS condition the generation process on textual input \citep{jeong2021diff, chen2021wavegrad2, GradTTS}, and works like DiffGAN-TTS, FastDiff and ProDiff \citep{liu2022diffgan, huang2022fastdiff, huang2022prodiff} prioritize generation efficiency and expressiveness.

Beyond efficiency, researchers have explored DDMs for controllable and expressive TTS. PromptTTS \citep{guo2023prompttts} and NaturalSpeech~2 \citep{shen2023naturalspeech} employ text prompts and speech prompts, respectively, to control speech style and content. In both methods, the conditional denoiser must undergo a specialized training process. Other methods for controlling the vocal characteristics require large quantities of annotated samples~\cite{guo2023emodiff} or retraining~\cite{kim2022guided}. We propose a speech editing method that requires no additional data or training and can be applied to any frozen diffusion-based TTS model that incorporates a bottleneck.



In the image synthesis domain, \citet{kwon2022diffusion} recently discovered a semantically meaningful latent space, named \textit{h-space}, providing versatile semantic editing capabilities. This discovery was further explored by \citet{Michaeli}, who proposed methods for identifying semantic directions. To the best of our knowledge, despite the widespread adoption of diffusion models for TTS in recent years, the existence of a hidden semantic space has not been examined in the speech synthesis domain. This raises intriguing questions regarding the possibility of facilitating latent space arithmetics for audio editing.

In this work we investigate the existence of a semantic space within diffusion-based TTS systems. We study the properties of \textit{h-space} in pre-trained TTS models and uncover its acoustically-semantic characteristics. Then, we propose novel methods for semantic speech editing through both supervised and unsupervised latent space arithmetics, inspired by \citet{Michaeli} and adapted to the speech synthesis domain for the first time. Our work offers intuitive and efficient audio editing techniques that require neither classifier guidance~\cite{guo2023emodiff}, model retraining~\cite{kim2022guided}, optimization, speech prompts nor any architecture modifications. To validate our methods, we present extensive experiments that demonstrate effective and high-quality edited speech synthesis. 

\section{Methods}
\label{sec:methods}
\subsection{Denoising Diffusion Models}
DDMs generate realistic data by iteratively removing noise, and are applicable to various modalities like images, audio, and text \citep{ho2020denoising}. 
Initially formulated as Markov chains, DDMs can be unified under stochastic differential equations (SDEs) \citep{song2020score} and adapted for TTS \cite{GradTTS}.
DDMs consist of two processes: forward diffusion and reverse diffusion. The forward process transforms any data distribution to a Gaussian $\mathcal{N}(\boldsymbol{\mu},\boldsymbol{\Sigma})$ via an SDE.
The reverse diffusion process is defined by another SDE:
\begin{equation*}
    d\rvx_t=\frac{\beta_t}{2}\left(\boldsymbol{\Sigma}^{-1}(\boldsymbol{\mu}-\rvx_t)-s(\rvx_t)\right)dt
+\sqrt{\beta_t}d\rvw_t
\end{equation*}
where $w_t$ is a Brownian motion, $\beta_t$ is a predefined noise schedule, and $s(\rvx_t)=\nabla\log p_t(\rvx_t)$ is the score function of the probability density function $p_t$ of $\rvx_t$.
The reverse process is typically solved via the Euler-Maruyama scheme \citep{kloeden1992stochastic}, discretizing the time interval $[0, 1]$ into $T$ time-steps.
By training a denoising neural network $s_t^\theta(\rvx_t)\approx s(\rvx_t)$ to estimate the true score function, we can sample from the target data distribution. Within TTS systems, DDMs are utilized as acoustic models, vocoders, or as end-to-end solutions.

\subsection{Semantic Audio Editing via Latent Space Manipulation}
\label{sec:edit}
\begin{figure}
    \centering
    \includegraphics[trim={1.5cm 2.8cm 1.5cm 1.5cm},clip, width=0.48\textwidth]{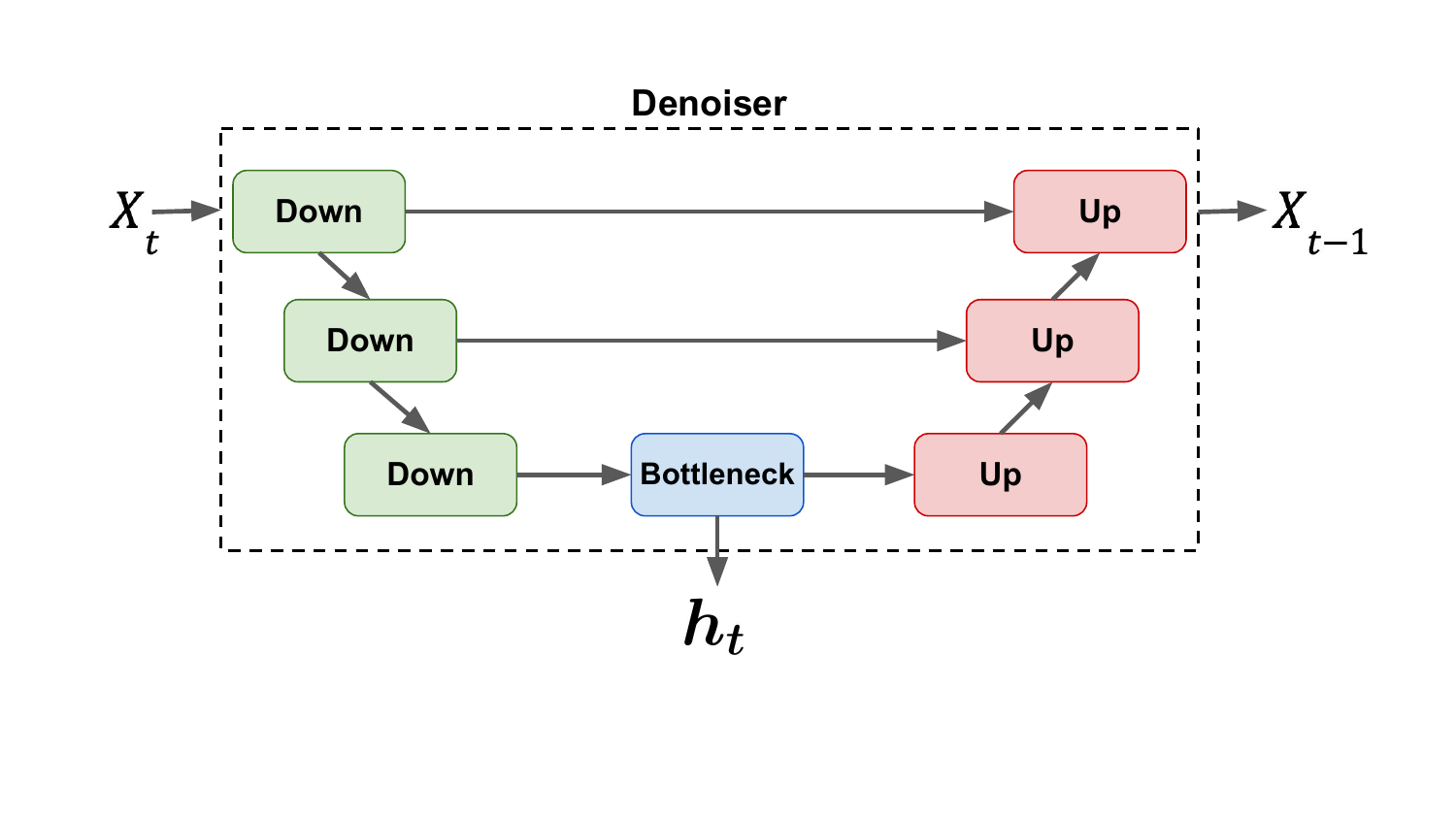}
    \caption{The \textit{h-space} of a diffusion model is defined as the concatenation of the bottleneck activations of the U-Net architecture.}
    \label{fig:hspace}
\end{figure}

We aim to discover a semantic latent space within frozen diffusion-based TTS models.
 We build upon the work of \citet{kwon2022diffusion} who introduced a semantic latent space in image diffusion models. 
Leveraging the standard implementation of the denoising network, $s_t^\theta(\cdot)$, as a U-Net architecture ~\citep{ronneberger2015u} in state-of-the-art  models, \citet{kwon2022diffusion}  examined the deepest feature maps, residing at the bottleneck of the network (visualized in Figure~\ref{fig:hspace}). These features are subsequently concatenated across all $T$ time-steps to construct the following latent code: 
\begin{equation}
    \rvh\triangleq \rvh_{T:1}=\texttt{concat}(\rvh_T,\rvh_{T-1},\dots,\rvh_1)
\end{equation}
This approach yields the \textit{h-space}: a latent space exhibiting favorable properties for versatile semantic editing and quality enhancement of images \citep{kwon2022diffusion, Michaeli}.

We adapt the concept of \textit{h-space} to the domain of TTS, demonstrating it encapsulates semantic information and performing semantic editing of synthesized speech through simple latent space arithmetics. Specifically, given a speech sample whose features are $\rvh\triangleq \rvh_{T:1}$ and a direction $\rvv\triangleq \rvv_{T:1}$, associated with desired acoustic attributes, we propose the following editing process:
\begin{equation}
    \rvh^{edit}\triangleq \rvh^{edit}_{T:1} = \rvh_{T:1}+\lambda\cdot \rvv_{T:1}
    \label{eq:edit}
\end{equation}
where $\lambda$ controls edit intensity, and both addition and scaling are element-wise.
Replacing the latent code $\rvh$ with $\rvh^{edit}$ during the generation process embodies the synthesized speech with the acoustic attributes related to the chosen editing direction. 

\begin{figure*}
    \centering
    \includegraphics[clip, trim=0.7 5.55cm 0.7 0.25cm , width=0.98\textwidth]{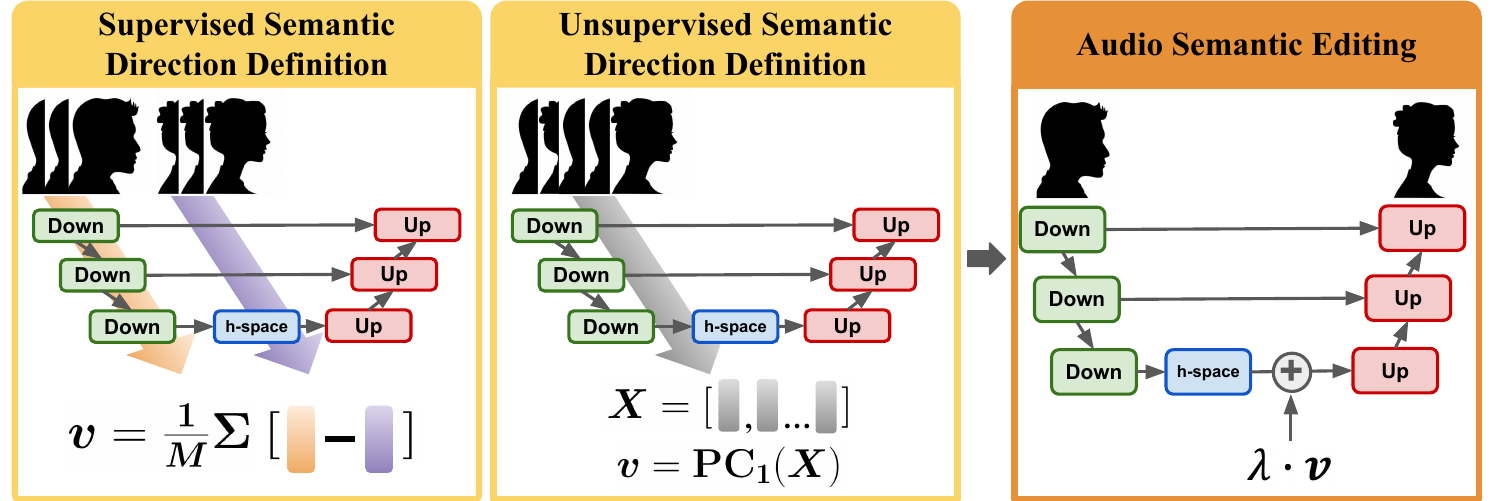}
    \caption{We propose a simple yet effective semantic audio-editing method. A latent semantic direction is defined either in a supervised or an unsupervised manner, and the corresponding speech attribute is edited by applying that direction to the latent space during the generation process of a new speech sample. The method is demonstrated with the male-to-female editing direction.}
    \label{fig:pipeline}
\end{figure*}

Having established the editing framework, we next derive editing directions via the following (illustrated in Figure \ref{fig:pipeline}):

\noindent \textbf{Supervised Approach.}
Given a pre-trained TTS model and a specific text prompt, we generate \textit{m} paired samples $\{(\rvx_{(k)}^+, \rvx_{(k)}^-)\}_{k=1}^m$ characterized by the presence or absence of a desired attribute. Denoting their matching latent codes by $\{(\rvh_{(k)}^+,\rvh_{(k)}^-)\}_{k=1}^m$, we define a semantic direction towards this attribute as
\begin{equation}
    \rvv\triangleq \Delta \rvh = \frac{1}{m} \sum_{k=1}^m (\rvh_{(k)}^+ - \rvh_{(k)}^-)
    \label{eq:supervised}
\end{equation}
\noindent \textbf{Unsupervised Approach.}
For a given text input, we generate speech samples and extract their bottleneck features $\{\rvh_t^{(i)}\}_{i=1}^n$ for each time-step $t\in[1,T]$. Applying PCA per time-step, we define the editing direction $\rvv^{(j)}$ as a concatenation of the $j$th principal components across time-steps. Surprisingly, the main principle components display clear semantic attributes as gender and intensity. The above framework unlocks semantic editing in diffusion-based TTS models, facilitating expressive and diverse speech synthesis.

\section{Experimental Results}
\subsection{Implementation Details}


For demonstration, we use Grad-TTS~\citep{GradTTS}, a recently published publicly available diffusion-based TTS model, trained on LibriTTS~\citep{LibriTTS}. However, our method can also be applied to any other unguided diffusion-based TTS model that contains a bottleneck. Grad-TTS takes a text and a speaker embedding as input, and generates a clean mel-spectogram through a U-Net-based denoiser.
We use 10 diffusion timesteps for mel-spectogram generation, as suggested by Grad-TTS authors, followed by the Universal HifiGan vocoder \citep{HifiGAN} for waveform generation. 


\subsection{Supervised Latent Space Editing}
\label{sec:supervised}
We begin our analysis by exploring the semantic-capturing capabilities of \textit{h-space} using the per-speaker gender annotations available for LibriTTS. Capturing the latent code during all timesteps of the generation process and following Equation~\ref{eq:supervised}, we calculate the male-to-female latent direction, and utilize it for audio editing as outlined in Equation~\ref{eq:edit}. As the latent vectors' lengths vary with the input texts, editing direction is defined per text. 
For a comparable baseline, we use another, simpler, approach for gender-editing: manipulating the speaker embedding, which is provided to the model as an input. We calculate the male-to-female direction in the speaker embedding space in a similar manner by averaging the differences of speaker embeddings between pairs of male and female speakers. The input speaker embedding is modified by adding this direction with different scales ($\lambda$). We provide supplemental samples, demonstrating the suggested audio editing methods: \href{https://latent-analysis-grad-tts.github.io/speech-samples/}{https://latent-analysis-grad-tts.github.io/speech-samples/}.

\begin{figure}
    \begin{subfigure}[b]{0.45\textwidth}
        \centering
        \includegraphics[width=\textwidth, trim=0 0.6cm 0 0.3cm , clip]{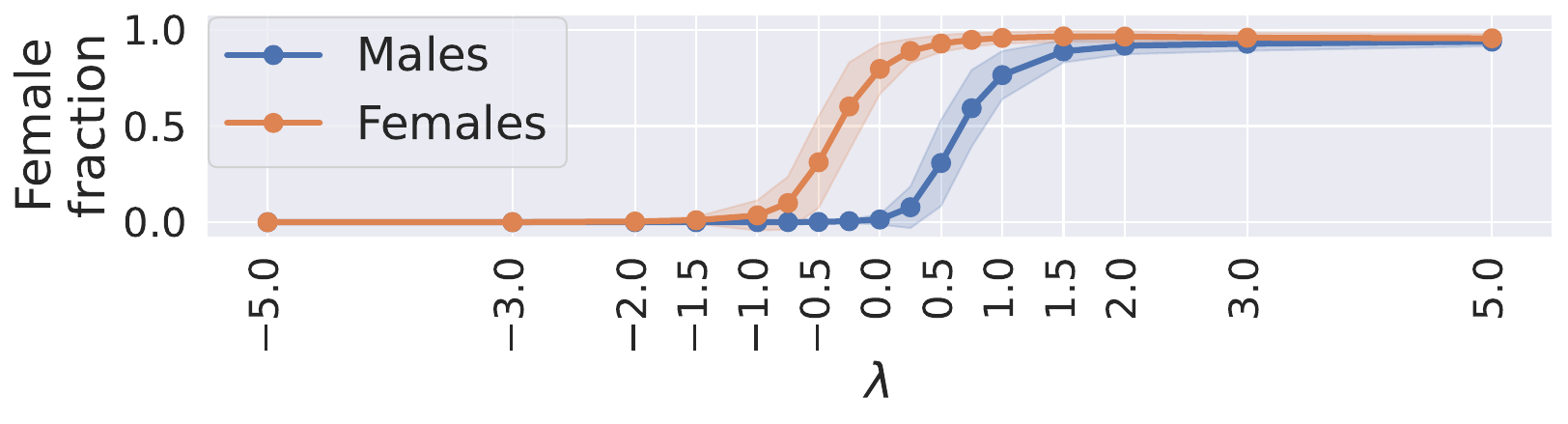}
        \caption{Latent space editing}
        \label{fig:latent_arithmetics}
    \end{subfigure}

    \begin{subfigure}[b]{0.45\textwidth}
        \centering
        \includegraphics[width=\textwidth, trim=0 0.6cm 0 0.3cm , clip]{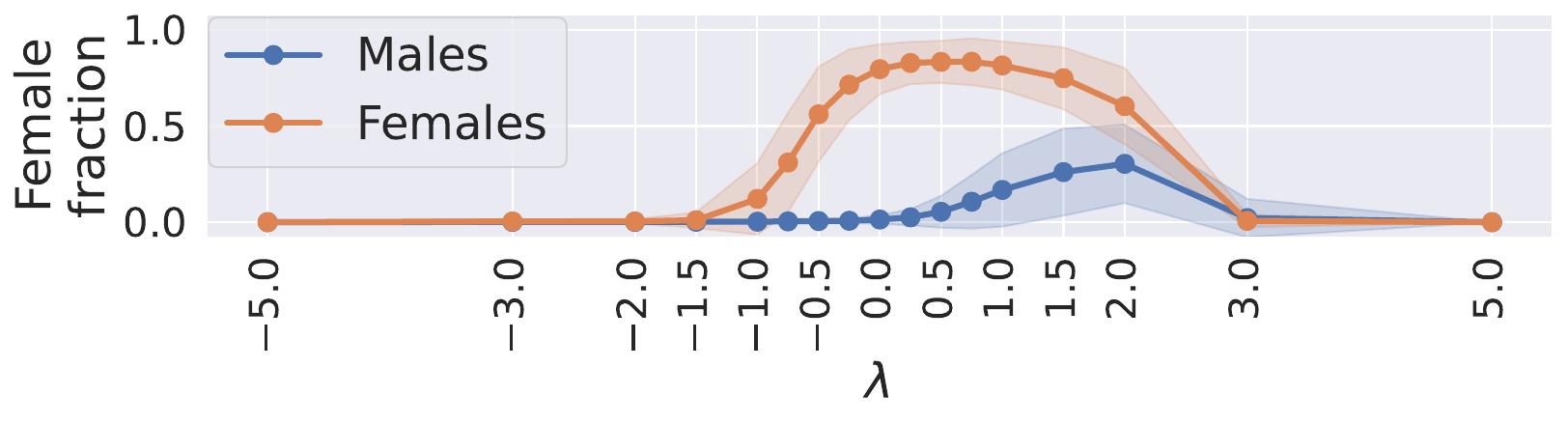}
        \caption{Speaker embedding editing}
        \label{fig:speaker_arithmetics}
    \end{subfigure}
    \caption{Supervised latent space editing allows gender manipulation, while speaker embedding editing does not. The percentage of samples classified as female is reported separately for male and female input spakers, averaged across 50 texts and all speakers (standard deviation, STD, is shaded).}
    \label{fig:supervised_fig}
\end{figure}

\textbf{Semantic properties evaluation}. We fine-tuned a speech gender classifier \citep{Classifier} on Grad-TTS outputs, acknowledging the different quality of synthesized speech compared to human-recorded samples. Then, we applied gender editing via both latent space and speaker embedding editing using varying $\lambda$ values, across the first 50 texts of the LibriTTS test set and all 247 speakers. In Figure~\ref{fig:supervised_fig} we report the fraction of samples classified as female for each $\lambda$ value, averaged across input male and female speakers separately. Latent space editing exhibits a monotonic behavior with more samples classified as female as $\lambda$ increases. On the contrary, speaker embedding editing fails to transform male voices to female ones, and when $\lambda \geq 3$ even originally female voices are not classified as such.

Additionally, 10 human evaluators classified speech samples as male or female. Analyzing samples from 20 different speakers, we compared the unedited Grad-TTS outputs to the gender-edited samples. For an effective gender alteration as shown in Figure \ref{fig:supervised_fig}, we used $\lambda=2$ and $-2$ for male-to-female and female-to-male editing, respectively. Table \ref{tab:supervised} presents the accuracy of predicting the expected gender (original gender for original samples, and contrasting gender for edited samples). Comparing to speaker editing, latent space editing achieves a classification accuracy that is higher by 24\%, with statistical significance (p-value $ < 0.001$).


\begin{table}[t]
    \centering
    \begin{tabular}{lcc}
    \hline
    \textbf{Method} & \textbf{Gender acc. $\mathbf{\uparrow}$} & \textbf{MOS $\mathbf{\uparrow}$} \\
    \hline
    {Grad-TTS} & {$0.82 \pm 0.14$} & {$3.95 \pm 0.15$}  \\
    {Speaker Editing} & \tikzmark{SA}{$0.76 \pm 0.24$} & \tikzmark{SA2}{$3.19 \pm 0.17$} \\
    {Latent Editing} & \tikzmark{LA}{$0.94 \pm 0.07$} & \tikzmark{LA2}{$3.59 \pm 0.24$} \\
    \hline
    
    \small 
    \textbf{*** p-value < 0.001} & { }

    \end{tabular}

    \begin{tikzpicture}[overlay,remember picture,black, thick]%
        \coordinate (LA Extended) at ($(LA.west)+(-1.3pt,0ex)$);
        \coordinate (SA Extended)  at ($(SA.west) +(-1.3pt,0ex)$);
        \draw (LA.west) -- (LA Extended) 
            -- (SA Extended) -- (SA.west);
        \node [left, AST style, scale=0.45] at ($(LA.west)+(-1.pt,1.35ex)$) {\ThreeAst};
    \end{tikzpicture}
    \begin{tikzpicture}[overlay,remember picture,black, thick]%
        \coordinate (LA2 Extended) at ($(LA2.west)+(-1.3pt,0ex)$);
        \coordinate (SA2 Extended)  at ($(SA2.west) +(-1.3pt,0ex)$);
        \draw (LA2.west) -- (LA2 Extended) 
            -- (SA2 Extended) -- (SA2.west);
        \node [left, AST style, scale=0.45] at ($(LA2.west)+(-1.pt,1.35ex)$) {\ThreeAst};
    \end{tikzpicture}

    \caption{Supervised latent space editing generates intelligible samples where the perceived speaker's gender is correctly classified, while speaker embedding editing does not. Average gender accuracy and MOS (mean $\pm$ STD) are reported. Latent-editing results compared to speaker-editing results are statistically significant (using ~\citet{Wilcoxon} rank sum test).}
    \label{tab:supervised}
\end{table}



\textbf{Acoustic properties evaluation}. To assess the perceived naturalness of the generated speech we measure the Mean Opinion Score (MOS), as quantified by 10 experienced evaluators on a scale of 1 to 5, across the same set of samples reported before. 
Table~\ref{tab:supervised} shows that the perceived naturalness of latent space editing, compared to speaker editing, is higher by $12\%$, a statistically significant difference (p-value $<0.001$).
This, combined with the superior perceived gender editing quality, reinforces the latent space's capability to encapsulate non-trivial semantic information.

\subsection{Unsupervised Latent Space Editing}
Next, we investigate semantically meaningful directions in \textit{h-space} without prior annotations. First, we generated speech samples for the first 50 test texts of LibriTTS and across all 247 speakers, and recorded the latent vectors $\textbf{h}_{T:1}$. Then, following the unsupervised process defined in Section~\ref{sec:edit}, PCA of the latent space was performed for each text across all samples, calculating the first 3 principal components (PCs). As vocal attributes, for each speech sample we extracted its speaker's gender from the metadata, and measured its intensity, Harmonics-to-Noise Ratio (HNR), and pitch using the Parselmouth Python package~\citep{parselmouth}.

\begin{figure}[t]
    \centering
    \includegraphics[width=0.45\textwidth, trim=0 0.3cm 0 0.75cm , clip]{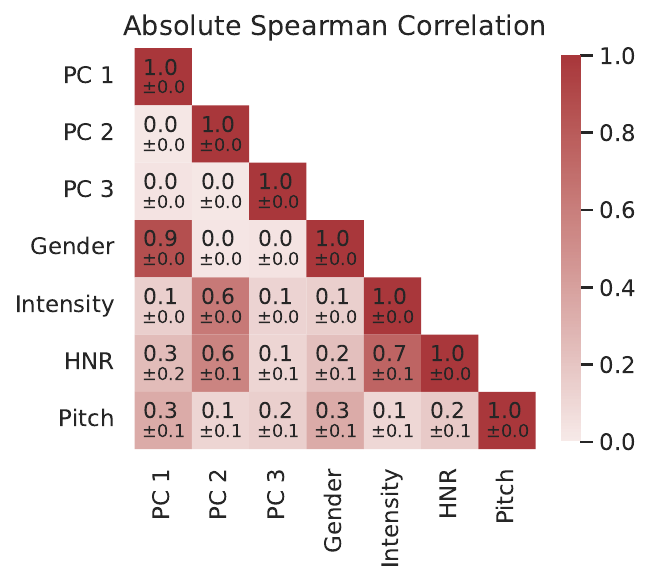}
    \caption{Absolute values of the Spearman correlation between the latent space PC-projections and the vocal attributes of the generated speech. We report mean and STD across all speakers, timesteps, and 50 texts.}
    \label{fig:correlation_fig}
\end{figure}

The latent vectors of each sample were projected onto each PC. Next, we calculated the absolute value Spearman correlation between each vocal attribute and  PC-projection vector, averaging across texts and timesteps. As Figure~\ref{fig:correlation_fig} shows, PC1 strongly correlates ($\rho=0.9\pm0.0$) with speaker's gender (also see Figure \ref{fig:pcs} in Appendix \ref{sec:appendix_results}), while PC2 correlates ($\rho=0.6\pm0.1$) with intensity and HNR. Other PCs and vocal attributes show no significant correlation and neither did random projections in the latent space (see Figure \ref{fig:correlation_fig_rand_proj} in Appendix \ref{sec:appendix_results}).

\begin{figure}
    \begin{subfigure}[b]{0.45\textwidth}
        \centering
        \includegraphics[width=1\textwidth, trim = 0 0.6cm 0 0.3cm, clip]{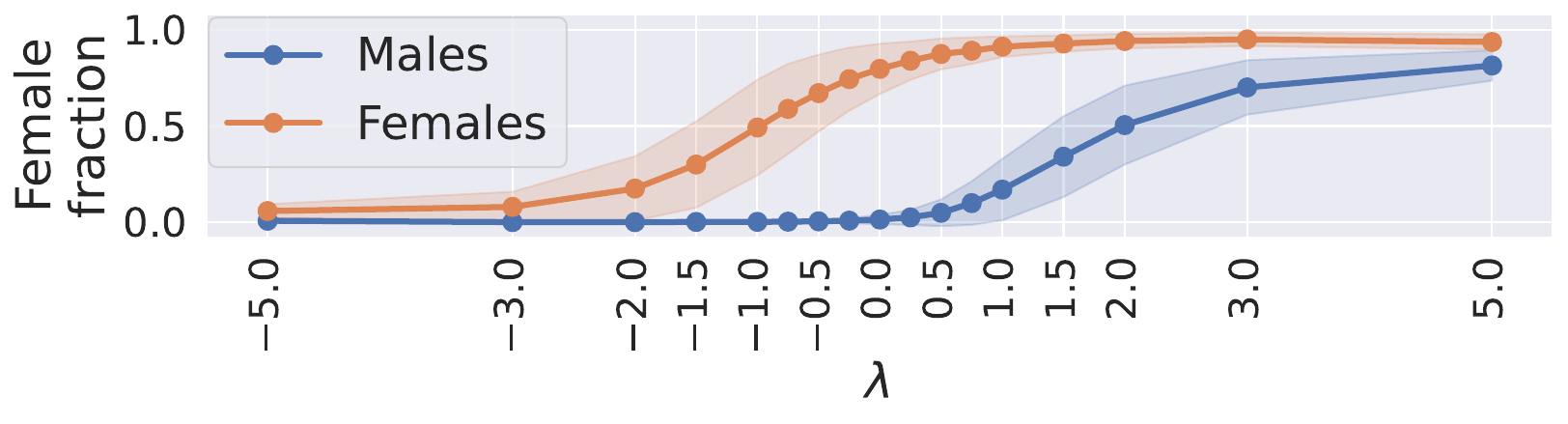}
        \caption{Latent editing along PC1}
	\label{fig:PC0}
    \end{subfigure}

    \begin{subfigure}[b]{0.45\textwidth}
        \centering
        \includegraphics[width=1\textwidth, trim = 0 0.6cm 0 0.3cm, clip]{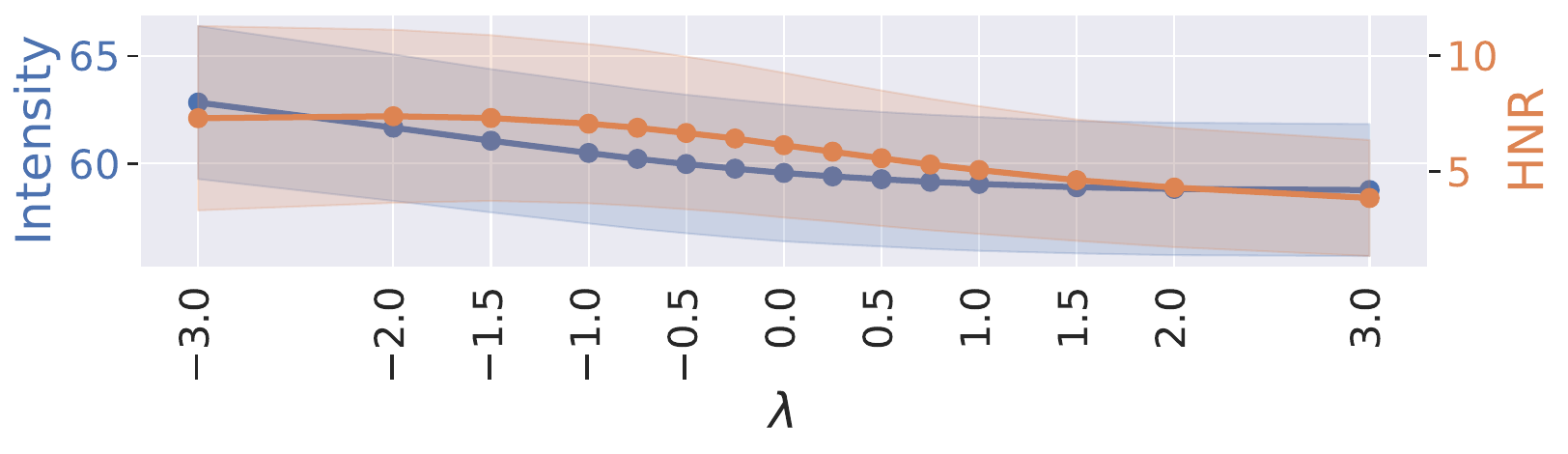}
        \caption{Latent editing along PC2}
	\label{fig:PC1}
    \end{subfigure}
\caption{Interpolation along the semantic directions revealed by PCA changes the vocal attributes accordingly. The reported values are averaged over 50 texts and all speakers. Shaded area is the STD.}
\label{fig:unsup}
\end{figure}

\textbf{Semantic properties evaluation}. Using PCs as editing directions in \textit{h-space}, we explore speech editing capabilities. Since the PCs are unitary vectors, the editing directions were normalized to the norm of the latent vectors. Intriguingly, our experiments indicate that decreasing the editing norm at later timesteps improves acoustic quality. 
As can be seen in Figure~\ref{fig:PC0}, interpolation along PC1 exhibits a smooth transition between male and female voices. 
Simialrly, intensity and HNR decrease when interpolating along PC2 (see Figure \ref{fig:PC1}).
Importantly, no gender-editing occurs when interpolating along PC2 (see Figure \ref{fig:disentanglement} in Appendix \ref{sec:appendix_results}).

Additionally, we measured the accuracy of gender classification as evaluated by human annotators on the same 20 speakers. Following the analysis in Figure \ref{fig:unsup}, to ensure effective gender alteration, we used $\lambda=3$ or $-3$ for originally male or originally female speakers, respectively, while editing along PC1. For PC2, $\lambda=-2$ was used to maximize HNR. PC1-edited samples were successfully classified as the contrasting gender with an even higher accuracy than un-edited ones (Table \ref{tab:unsuervised}).


\begin{table}
    \centering
    \begin{tabular}{lcc}
    \hline
    \textbf{Method} & \textbf{Gender acc. $\mathbf{\uparrow}$} & \textbf{MOS $\mathbf{\uparrow}$}\\
    \hline
    {Grad-TTS} & \tikzmark{Orig}{$0.82 \pm 0.14$} & {$3.95 \pm 0.15$} \\
    {PC1 Editing} & \tikzmark{PC1}{$0.88 \pm 0.14$} & {$3.86 \pm 0.20$} \\
    {PC2 Editing} & \tikzmark{PC2}{$0.82 \pm 0.16$} & {$3.98 \pm 0.17$} \\
    \hline
   
    \small 
    \textbf{* p-value < 0.05} & { }

    \end{tabular}

    \begin{tikzpicture}[overlay,remember picture,black, thick]%
        \coordinate (PC1 Extended) at ($(PC1.west)+(-1.75pt,0ex)$);
        \coordinate (Orig Extended)  at ($(Orig.west) +(-1.75pt,0ex)$);
        \draw (PC1.west) -- (PC1 Extended) 
            -- (Orig Extended) -- (Orig.west);
        \node [left, AST style, scale=0.75] at ($(PC1.west)+(-1.pt,1.35ex)$) {\OneAst};
    \end{tikzpicture}
    \caption{Gender accuracy and MOS results (mean $\pm$ STD) for unsupervised latent space editing.} 
    \label{tab:unsuervised}
\end{table}

\textbf{Acoustic properties evaluation}. Using the same setup, we assessed speech naturalness using MOS. Table \ref{tab:unsuervised} compares the perceived naturalness of samples with and without latent editing, presenting similar scores between the groups. 
The Wilcoxon rank sum test indicated no statistically significant difference in the MOS between groups (p-value $\gg 0.05$).
Thus, we conclude that speech editing through unsupervised latent space manipulation does not compromise the acoustic quality.

\section{Conclusions}
In this paper, we identify the semantic properties of the latent space of diffusion-based TTS models, referred to as \textit{h-space}. We develop supervised and unsupervised methods for finding interpretable directions in that space, and provide empirical qualitative evidence for their semantic quality. Moreover, the proposed latent space editing methods preserve and even enhance the acoustic quality of the generated samples. This study presents evidence regarding specific vocal attribute manipulation, such as gender or intensity. However, the presented method can be applied to any vocal attribute present in the data. 

\clearpage
\section*{Limitations and Ethics}
This study is subject to several limitations. We demonstrated our analysis on the Grad-TTS model \cite{GradTTS} (trained on LibriTTS dataset \citep{LibriTTS}), and used the Universal HifiGAN \citep{HifiGAN} for waveform generation. These are all publicly available for our research purposes. We do not develop novel TTS models from scratch, and focus on analysing existing ones. Under these settings, several limitations apply to our analysis:

\begin{enumerate}
    \item LibriTTS is an English-only dataset, hence other languages are not supported by Grad-TTS, and were not analyzed.
    \item LibriTTS is an audio-book reading dataset, and besides the speaker's gender no vocal attributes are provided. Therefore, we were limited to use the speaker's gender and the statistical audio attributes that we measure directly from the waveform. Properties such as emotion could not be analysed under these settings. We only refer to "male" or "female" voices to align with the original metadata.
    \item Our method is general and can be applied to any frozen unguided diffusion-based TTS model that contains a bottleneck. However, since we were limited to publicly available models, we chose to focus on analysing the Grad-TTS model.
    \item The acoustic quality of generated samples is bounded by the quality of the TTS system, including the Grad-TTS spectogram denoiser and the Universal HifiGAN vocoder quality.
    \item The system cannot generate speech with a custom voice, as it does not take a voice-prompt as input. Thus, our edited audios are limited to the given subspace of speaker voices. This also points to the fact that our work does not pose risks regarding deep-fake or identity theft.
    
\end{enumerate}

\section*{Acknowledgements}
We thank Michael Hassid for the great feedback and moral support.

\clearpage
\bibliography{anthology,custom}

\appendix

\clearpage
\section{Additional Results}
\label{sec:appendix_results}
Further results supporting our main claims are presented in the following section.

\begin{figure}[hb]
    \centering
    \includegraphics[trim={0.4cm 0.5cm 0.4cm 0.5cm}, clip, width=0.5\textwidth]{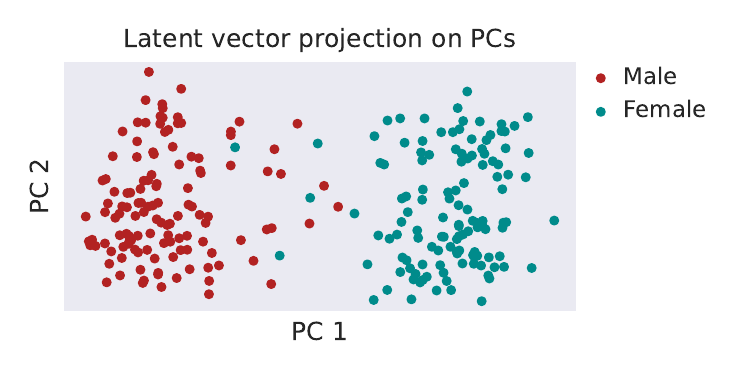}
    \captionof{figure}{PC1 separates male from female speakers. Shown are the projection of latent spaces of samples generated with male and female speaker IDs onto PC1 and PC2.}
    \label{fig:pcs}
\end{figure}

An analysis of the PC1 and PC2 components of all the male and female speakers from LibriTTS is shown in Figure~\ref{fig:pcs}. It can be seen that PC1 provides an excellent separation between male and female voices. In contrast, PC2 does not provide such a separation. 

\begin{figure}[hb]
    \centering
    \includegraphics[width=0.5\textwidth]{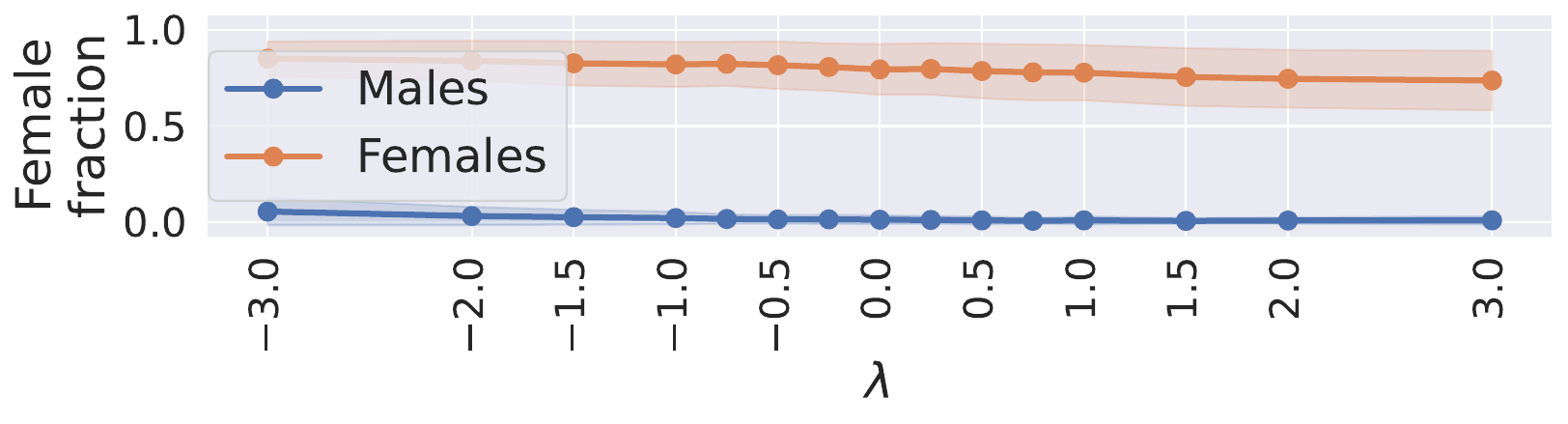}
    \captionof{figure}{Interpolation along PC2 does not edit the perceived speaker's gender, indicating disentanglement of editing directions.}
    \label{fig:disentanglement}
\end{figure}

Figure~\ref{fig:disentanglement} presents the interpolation across PC2 for different $\lambda$ values while monitoring the perceived speaker's gender. In line with expectations, interpolating across this editing direction does not affect the perceived speaker's gender, and it remains relatively unchanged. This is another indication of the disentanglement between the different editing directions found in the latent space by using our method.

\begin{figure*}[ht]
    \centering
    \includegraphics[width=0.85\textwidth]{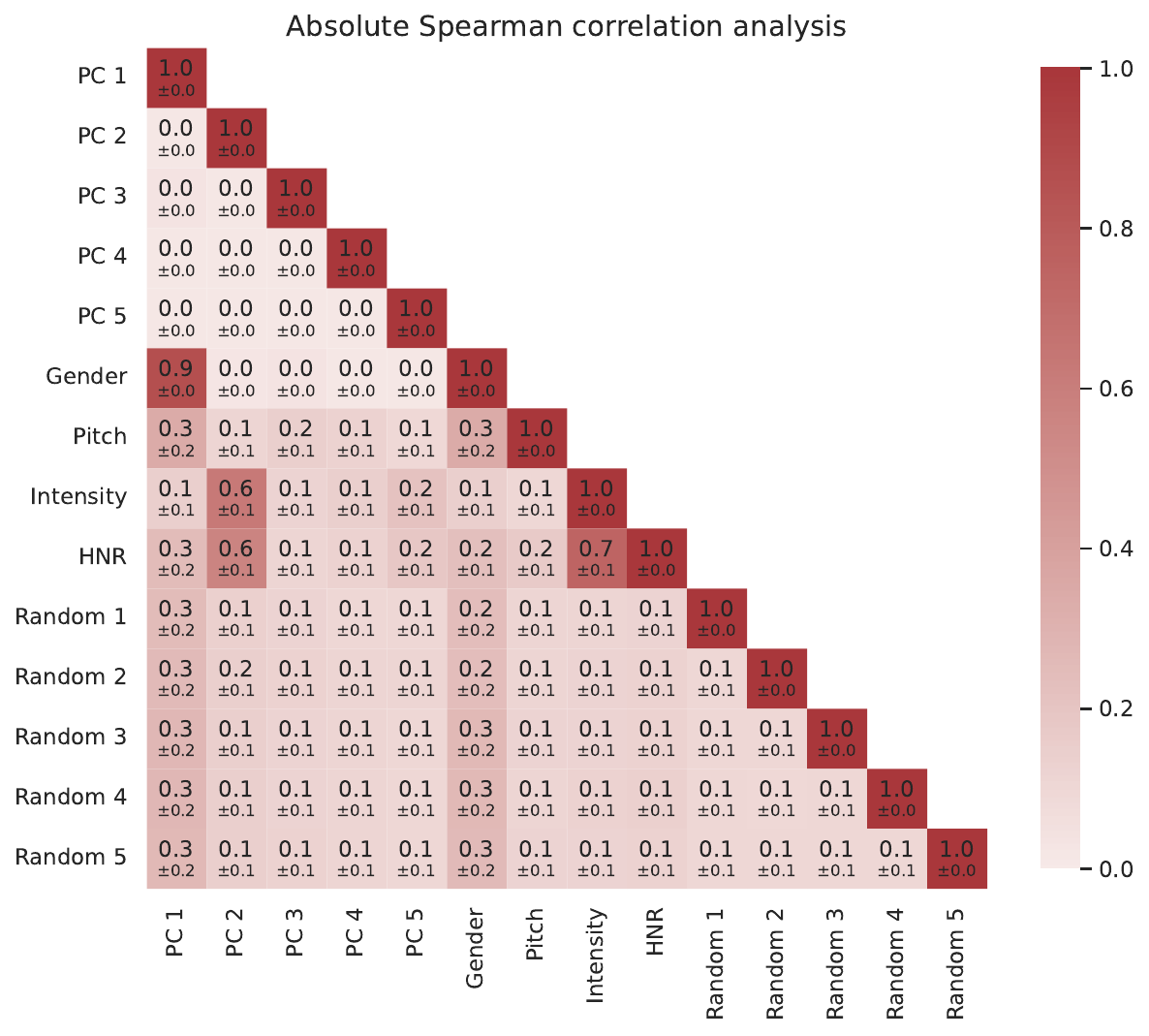}
    \captionof{figure}{Principal components of latent space correlate with attributes of the generated audio. Shown are the mean and STD of the absolute value Spearman correlation of the PCs of the latent space, vocal attributes of the generated audios, and random projections of the latent space, averaged across all speakers, timesteps and 50 texts.}
    \label{fig:correlation_fig_rand_proj}
\end{figure*}

A more detailed version of Figure~\ref{fig:correlation_fig} is presented in Figure~\ref{fig:correlation_fig_rand_proj}, with random latent space projections and additional PC directions. As can be seen, only PC1 and PC2 exhibit significant correlations with the vocal attributes that were tested. Contrary to PCs, random projections do not correlate with any vocal attribute. This observation supports our claim that the latent space is capturing unique semantic properties.

\section{Human Annotators}
\label{sec:appendix_mos}
\subsection{Human Evaluation of Perceived Speaker's Gender}
To evaluate the perceived speaker's gender of generated samples, we used the Amazon Mechanical Turk (MTurk) crowd-sourcing platform. The MTurk workers we recruited and filtered had an approval rate above 50\% and were located in the USA. The workers were instructed to classify the gender of each sample (binary classification). Each crowd worker was given the following instruction:
"You are given an audio sample generated from a Text-To-Speech computer program. To the best of your ability, please classify the gender of the speaker in each audio sample. For better results, wear headphones and work in a quiet environment". We paid 0.02\$ per Human Intelligence Task (HIT), and each worker was paid 4\$ on average.

\subsection{Mean Opinion Score Evaluation}
To evaluate the quality of the generated speech, we utilized an internal annotation system. 34 experienced workers from the USA, who are native English speakers, have been assigned to assess the Mean Opinion Score (MOS) of the generated speech. Each worker was paid 0.34\$ per-task (annotating a 3-second audio file) and each worker was paid an average of 51\$ in total. The workers have been instructed to rate each speech sample quality based on the acceptable 5-point MOS score, Table \ref{tab:mos_scores} provides details regarding the scoring methodology used.

\begin{table}[h]
    \centering
    \footnotesize
    \begin{tabular}{cc}
        \toprule
        Score & Quality \\
        \midrule
        5.0 & Excellent (Completely defined) \\
        4.5 & \\
        4.0 & Good (Mostly defined) \\
        3.5 &  \\
        3.0 & Fair (Equally defined and undefined) \\
        2.5 &  \\
        2.0 &  Poor (Mostly undefined) \\
        1.5 &  \\
        1.0 & Bad  (Completely undefined) \\
        
        \bottomrule
    \end{tabular}
    \caption{Mean Opinion Score (MOS) scoring schema.}
    \label{tab:mos_scores}
\end{table}

\end{document}